\documentclass[12pt,english]{article}
\usepackage{a4wide}
\usepackage{array}
\usepackage{subfigure}
\usepackage{graphicx}

\makeatletter

\newcommand{\noun}[1]{\textbf{#1}}

\providecommand{\tabularnewline}{\\}

\makeatletter

\@addtoreset{equation}{section}

\@addtoreset{figure}{section}

\@addtoreset{table}{section}
\makeatother

\usepackage{babel}
\makeatother
\begin{document}

\title{Double sine-Gordon model revisited}

\author{G. Tak\'acs%
\footnote{Email: takacs@elte.hu%
} and F. W\'agner%
\footnote{E-mail: wferi@niif.hu%
}\\
\\
Theoretical Physics Research Group, Hungarian Academy of Sciences\\
H-1117 Budapest, P\'azm\'any P\'eter s\'et\'any 1/A}

\maketitle
\begin{abstract}
We reconsider the mass spectrum of double sine-Gordon theory where
recent semiclassical results called into question the previously accepted
picture. We use the Truncated Conformal Space Approach (TCSA) to investigate
the claims. We demonstrate that the numerics supports the original
results, and strongly disagrees with those obtained from semiclassical
soliton form factor techniques. Besides the numerical analysis, we
also discuss the underlying theoretical arguments.
\end{abstract}

\section{Introduction}

Double sine-Gordon theory has attracted interest recently chiefly
because it is a prototype of non-integrable field theory which can
be understood by application of techniques developed in the context
of integrable field theories \cite{delfino_mussardo}. It has several
interesting applications such as to the study of massive Schwinger
model (two-dimensional quantum electrodynamics) and a generalized
Ashkin-Teller model (a quantum spin system) which are discussed in
\cite{delfino_mussardo}. Another application to the one-dimensional
Hubbard model is examined in \cite{nersesyan} (together with the
generalized Ashkin-Teller model mentioned above). A further potentially
interesting application of the two-(and multi-)frequency sine-Gordon
model is for ultra-short optical pulses propagating in resonant degenerate
medium \cite{bullough}. 

In an earlier work the authors (and coworkers) studied this model
using non-perturbative finite size techniques and established the
phase diagram of the model \cite{dsg} (which was recently extended
to the multi-frequency generalization \cite{tgzs}). In particular
we used form factor perturbation theory introduced by Delfino and
Mussardo in \cite{delfino_mussardo} to predict mass spectra and verify
them using the truncated conformal space approach (TCSA) \cite{yurov_zamolodchikov}. 

Recently Mussardo et al. \cite{semicl} applied a semiclassical soliton
form factor technique developed by Goldstone and Jackiw \cite{goldstone-jackiw}
to study the mass spectrum of the theory. They obtained results that
contradict explicitely some of the results obtained using form factor
perturbation theory in \cite{dsg}. In principle, the truncated conformal
space approach applied in \cite{dsg} could decide the question, but
the accuracy achieved there is not sufficient for this purpose.

The present work reports a systematic study of this issue using improved
TCSA numerics which do have the required accuracy. Besides the numerical
comparisons we also address the theoretical arguments raised in \cite{semicl}
to support their case. We find a convincing and consistent picture,
both numerically and theoretically, which allows us to clarify the
issue. 

We would like to stress that the problem at hand is much more than
a simple numerical discrepancy: it is the validity of the form factor
perturbation theory and that of semiclassical techniques that is at
stake here, both of which are very promising tools for non-perturbative
investigation of non-integrable quantum field theories.

We start with recalling some necessary facts about the double sine-Gordon
model to set the stage in Section 2. Section 3 describes the mass
spectrum obtained from form factor perturbation theory in \cite{dsg}
on the one hand and from the semiclassical soliton form factor technique
on the other \cite{semicl}. Section 4 describes the numerical work
and the results obtained, while Section 5 is devoted to the underlying
theoretical issues. We conclude briefly in Section 6.

\section{The double sine-Gordon model}

The classical model is defined by the Lagrangian density\[
\mathcal{L}_{\mathrm{classical}}=\frac{1}{2}\partial_{\mu}\varphi\partial^{\mu}\varphi-\mu_{0}\cos\beta\varphi-\lambda_{0}\cos\left(\frac{\beta}{2}\varphi+\delta\right)\]
For $\lambda_{0}=0$ we recover standard sine-Gordon theory, where
the spectrum is known to consist of a soliton doublet of mass \[
M_{0}=\frac{8\sqrt{\mu_{0}}}{\beta}\]
and a continuum of breathers\[
m=2M_{0}\sin\frac{\pi}{2}\alpha\qquad,\qquad0<\alpha<1\]
We define the quantum version of the model using perturbed conformal
field theory as in \cite{dsg}. The Hamiltonian can be written\[
H_{\mathrm{PCFT}}=H_{\mathrm{c=1}}+\mu\int dx\::\cos\beta\varphi:+\lambda\int dx\::\cos\left(\frac{\beta}{2}\varphi+\delta\right):\]
where \[
H_{\mathrm{c=1}}=\int dx\::\frac{1}{2}\left(\partial_{t}\varphi\right)^{2}+\frac{1}{2}\left(\partial_{x}\varphi\right)^{2}:\]
is the Hamiltonian of a $c=1$ free boson CFT. The quantum spectrum
for $\lambda=0$ consists of a soliton doublet of mass $M$ which
is related to the coupling $\mu$ by \cite{mass_scale}\[
\mu=\frac{2\Gamma(\Delta)}{\pi\Gamma(1-\Delta)}\left(\frac{\sqrt{\pi}\Gamma\left(\frac{1}{2-2\Delta}\right)M}{2\Gamma\left(\frac{\Delta}{2-2\Delta}\right)}\right)^{2-2\Delta}\qquad,\qquad\Delta=\frac{\beta^{2}}{8\pi}\]
which satisfies the semiclassical asymptotics\[
\mu\sim\frac{\beta^{2}}{64}M^{2}\qquad,\qquad\beta\,\rightarrow\,0.\]
which means that $\mu$ can be identified with $\mu_{0}$ in that
limit (similar results hold for $\lambda$ and $\lambda_{0}$). In
the quantum theory, the breather spectrum is discrete and their numbers
depend on the frequency $\beta$: \begin{equation}
m_{n}=2M\sin\frac{\pi}{2}\frac{n\beta^{2}}{8\pi-\beta^{2}}\qquad,\qquad n=1,\dots,\left[\frac{8\pi}{\beta^{2}}\right]-1\label{eq:br_masses}\end{equation}

Double sine-Gordon model can be considered as a non-integrable perturbation
of the integrable sine-Gordon field theory \cite{delfino_mussardo}.
Because of the presence of the second cosine term, the field must
be taken to have the periodicity \[
\varphi\equiv\varphi+\frac{4\pi}{\beta}\]
which is called a $2$-folded model following \cite{kfold}. Therefore
at $\lambda=0$ the theory has two degenerate vacua corresponding
to the classical solutions \[
\varphi=0\qquad\mathrm{and}\qquad\frac{2\pi}{\beta}\]
As a result, each breather comes in two copies, and soliton states
must be labeled with the vacua between which they mediate, thereby
implementing a so-called 'kink' structure \cite{kfold}. In this paper
we treat $\lambda$ as the non-integrable perturbing coupling constant,
although it is equally possible to swap the roles of the cosine terms
in this respect. 

The most interesting case is $\delta=\frac{\pi}{2}$ when for small
$\lambda$ the two vacua remain degenerate until a critical value
is reached where the theory undergoes a phase transition. This was
shown at the classical level (mean field theory) by \cite{delfino_mussardo}
to be of second order in the Ising universality class, and it was
established that this holds even after taking quantum fluctuations
into account (non-perturbatively) in \cite{dsg}. This is the model
in which we are the most interested, although we shall also use the
$\delta=0$ case as a testing ground for our numerical calculations.

\section{Mass spectrum predictions}

\subsection{Form factor perturbation theory}

Form factor perturbation theory (FFPT) was introduced in \cite{nonintegrable}
to provide a way of expanding physical quantities in the non-integrable
perturbing coupling. It was applied to the double sine-Gordon model
in \cite{dsg}. For our present purposes, we only need the mass shift
of the first breather, which reads%
\footnote{We corrected some misprints in the published version of \cite{dsg}.%
}

\begin{equation}
\delta m_{1}^{(k)}=\frac{\lambda\mathcal{G}_{\beta/2}(\beta)\mathcal{N}}{M}\cos\left(\pi(k-1)+\delta\right)\exp\left\{ -\frac{1}{\pi}\int_{0}^{\pi p}dt\frac{t}{\sin t}\right\} \label{eq:ffpt_b1}\end{equation}
where $k=1,2$ labels the two copies of the first breather, $p=\beta^{2}/\left(8\pi-\beta^{2}\right)$,
\begin{eqnarray}
\mathcal{G}_{a}(\beta) & = & \left[\frac{M\sqrt{\pi}\Gamma\left(\frac{4\pi}{8\pi-\beta^{2}}\right)}{2\Gamma\left(\frac{\beta^{2}/2}{8\pi-\beta^{2}}\right)}\right]^{\frac{a^{2}}{4\pi}}\nonumber \\
 & \times & \exp\left\{ \int_{0}^{\infty}\frac{dt}{t}\left[\frac{\sinh^{2}\left(\frac{a\beta}{4\pi}t\right)}{2\sinh\left(\frac{\beta^{2}}{8\pi}t\right)\sinh\left(t\right)\cosh\left(\left(1-\frac{\beta^{2}}{8\pi}\right)t\right)}-\frac{a^{2}}{4\pi}e^{-2t}\right]\right\} \label{lz_vev}\end{eqnarray}
and \begin{equation}
\mathcal{N}=\exp\left\{ 4\int\frac{dt}{t}\frac{\sinh(t)\sinh(pt)\sinh((1+p)t)}{\sinh^{2}2t}\right\} \label{Nfactor}\end{equation}
It is an important prediction of FFPT that the mass corrections vanish
when $\delta=\frac{\pi}{2}$ and that also \[
\delta m^{(1)}=-\delta m^{(2)}\]
in general.

\subsection{Masses from semiclassical soliton form factors}

In \cite{semicl} a semiclassical method, based on the Goldstone-Jackiw
semiclassical kink form factor \cite{goldstone-jackiw}, was used
to determine the spectrum of double sine-Gordon theory. Here we give
a brief summary of their results for the case $\delta=\frac{\pi}{2}$.
They claim that the perturbation by $\cos\left(\frac{\beta}{2}\varphi+\frac{\pi}{2}\right)$
induces a linear mass correction to the mass of the $n$th breather,
and obtain the following result:\begin{equation}
m_{n}^{(L,S)}=m_{n}\pm2\pi\frac{\lambda_{0}\beta}{\sqrt{\mu_{0}}}\left(\frac{1}{\beta^{2}}\sin\left(n\frac{\beta^{2}}{16}\right)-\frac{n}{16}\cos\left(n\frac{\beta^{2}}{16}\right)\right)+O\left(\lambda_{0}^{2}\right)\label{eq:mussardo_d2_orig}\end{equation}
where $L$ and $S$ index the splitting up of the $n$th breather
level into what they call a 'short' and a 'long' breather. 

The main problem and the central issue in this paper, is that this
disagrees with the FFPT results in the previous subsection that give
a vanishing linear correction to the breather masses, and it also
disagrees with the numerical studies in \cite{dsg} where the masses
of the two copies of any breather levels remained degenerate as long
as $\lambda$ was lower than the critical value, where a second order
phase transition induced a mass gap between the ground states and
also lifted the degeneracy of the breather masses. However, the numerical
studies in \cite{dsg} were not aimed at studying the breather levels
and the data produced for that work, while still accessible, have
insufficient precision to decide whether a splitting of the order
of (\ref{eq:mussardo_d2_orig}) is actually present. In addition,
the authors of \cite{semicl} presented some arguments to support
the result (\ref{eq:mussardo_d2_orig}), to which we shall return
later. For the moment the issue is to decide which of the predictions
is borne out by the dynamics of the model.

Using the soliton mass $M_{0}=8\sqrt{\mu_{0}}/\beta$, the mass correction
can be written as\[
\frac{\delta m_{n}^{(L,S)}}{M_{0}}=\pm16\pi\frac{\lambda_{0}}{M_{0}^{2}}\left(\frac{1}{\beta^{2}}\sin\left(n\frac{\beta^{2}}{16}\right)-\frac{n}{16}\cos\left(n\frac{\beta^{2}}{16}\right)\right)=\frac{\pi n^{3}\beta^{4}}{768}\frac{\lambda_{0}}{M_{0}^{2}}+O\left(\beta^{6}\right)\]
In terms of the dimensionless coupling constant \[
\chi_{0}=\frac{\lambda_{0}}{M_{0}^{2}}\]
this can be written\begin{equation}
\frac{\delta m_{n}^{(L,S)}}{M_{0}}=\pm\frac{\pi n^{3}\beta^{4}}{768}\chi_{0}+O\left(\beta^{6}\right)\label{eq:mussardo_leading}\end{equation}
where we again absorbed the difference between the classical and quantum
mass/coupling relations into the higher order correction term. 

The dimensionless coupling of the quantum theory is defined by\[
\chi=\frac{\lambda}{M^{2-\frac{\beta^{2}}{16\pi^{2}}}}\]
(where the difference in the exponent of $M$ is just the anomalous
dimension of the perturbing operator). In the semiclassical limit
$\chi$ only differs from $\chi_{0}$ by terms higher order in $\beta^{2}$
as discussed in the previous section, and similarly for $M$ and $M_{0}$.
Therefore we can still use formula (\ref{eq:mussardo_leading}) to
leading order in $\beta$ in the quantum theory just replacing $M_{0}$
by $M$ and $\chi_{0}$ by $\chi$\[
\frac{\delta m_{n}^{(L,S)}}{M}=\pm\frac{\pi n^{3}\beta^{4}}{768}\chi+O\left(\beta^{6}\right)\]
For our purposes we are only interested in the first breather ($n=1$)
and the comparisons to numerical results will be done at the following
couplings:\begin{eqnarray}
\beta=\frac{\sqrt{4\pi}}{2.5} & : & \frac{\delta m_{1}^{(L,S)}}{M}=\pm0.0165367\chi+\dots\nonumber \\
\beta=\frac{\sqrt{4\pi}}{2.2} & : & \frac{\delta m_{1}^{(L,S)}}{M}=\pm0.0275751\chi+\dots\label{eq:mussardo_prediction}\end{eqnarray}

\section{TCSA analysis}

We now aim to get the mass spectrum solving the model directly using
the so-called truncated conformal space approach (TCSA) introduced
by Yurov and Zamolodchikov in \cite{yurov_zamolodchikov}. It was
extended to perturbations of $c=1$ free boson models in \cite{frt1}
and applied to the double sine-Gordon model in \cite{dsg}. 

The application of TCSA to perturbations of $c=1$ free boson CFT
is described in detail in \cite{dsg} and references therein, therefore
we do not go into the details here. We mention only that this method
gives the finite volume spectrum of the theory, from which the masses
can be extracted by locating the so-called scaling region in which
the mass gap function (the energy difference between the one-particle
state and the corresponding vacuum state) is approximately constant.
More precisely, we locate the volume range where the magnitude of
the numerical derivative of the mass gap function is smallest and
define the mass as the value the mass gap function takes there. The
Hilbert space can be split up into sectors differing in the value
of the winding number of the field $\varphi$ and the eigenvalue of
total spatial momentum, and the mass gap functions we are interested
in can be measured from the sector with no winding and zero momentum.

We improved on the precision of the TCSA significantly with respect
to \cite{dsg}. First, we chose smaller values for $\beta$ which
makes TCSA converge faster in the models at hand. This is also advantageous
since it means we are closer to the classical limit and so we expect
to be able to test the semiclassical prediction (\ref{eq:mussardo_d2_orig}).
Second, we pushed the cutoff much higher using more machine power,
and also by applying the parity symmetries to project the Hilbert
space into even and odd subspaces, using that the TCSA Hamiltonian
is block diagonal in the appropriate basis. This made possible treating
more than $18000$ states in reasonable computing time, and allowed
us to achieve the accuracy which makes it possible to judge the correctness
of the semiclassical prediction (\ref{eq:mussardo_d2_orig}).

\subsection{$\delta=0$: testing the numerics}

In this case the model has a symmetry \begin{equation}
\varphi\:\rightarrow\:-\varphi\label{eq:Z2_d0}\end{equation}
which can be used to project the TCSA Hilbert space onto an even and
an odd sector. This reduces the number of states and helps to achieve
higher values of the cutoff. In order to be on safe ground, the program
was run both with and without a projection for several values of $\beta$,
$\chi$ and the cutoff. The resulting spectra were always identical
(within the precision of floating point arithmetics), independently
of the values of the parameters and of the cutoff, which confirms
that the symmetry projection was correctly implemented. 

We also compared the $\chi=0$ to predictions of exact $S$ matrix
theory, which they matched perfectly within the truncation errors.
The vacua can be found in the even, while the one-particle states
in the odd sector. They can be matched against each other using the
fact that the bulk energy constant of the two vacua are different,
therefore the two vacua (and also the two corresponding one-particle
states) have a different slope in the scaling region. The redefinition
$\lambda\,\rightarrow\,-\lambda$ leaves the total spectrum invariant
(in both the odd and the even sector separately), but exchanges the
role of the two sets of states, therefore it is expected that\[
m_{1}^{(1,2)}\left(-\lambda\right)=m_{1}^{(2,1)}\left(\lambda\right)\]
 We choose the following value for the coupling \[
\beta=\frac{\sqrt{4\pi}}{2.5}=1.4179631\dots\]
For mass determination, the cutoff was placed at level $18$ resulting
in $9334$ states in the even and $9319$ states in the odd sector
respectively.

\begin{figure}
\begin{center}\subfigure[The two masses, denoted here by m1 and m2, from TCSA and FFPT]{\includegraphics{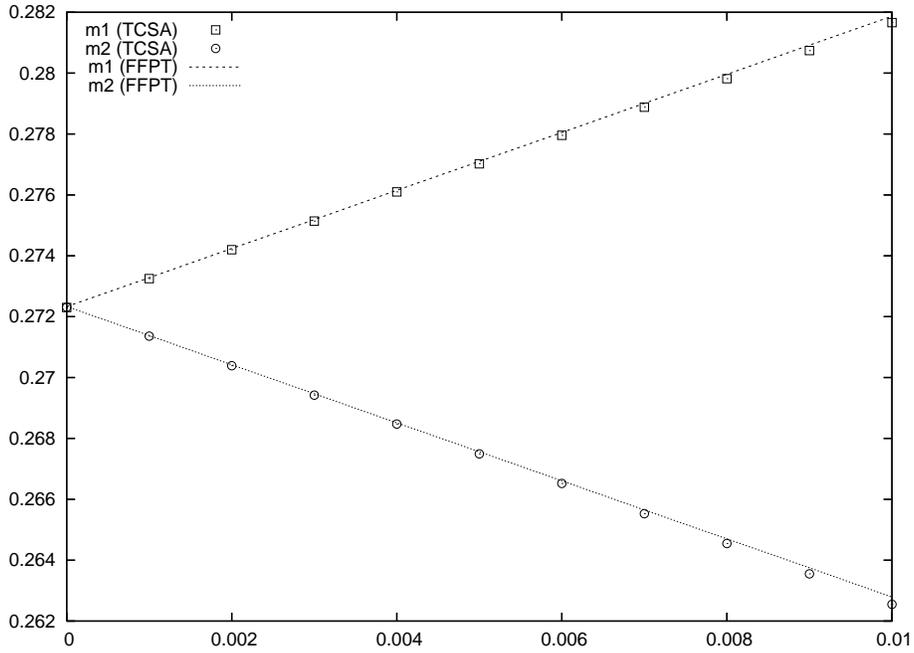}}\\
\subfigure[The mass splitting from TCSA and from FFPT]{\includegraphics{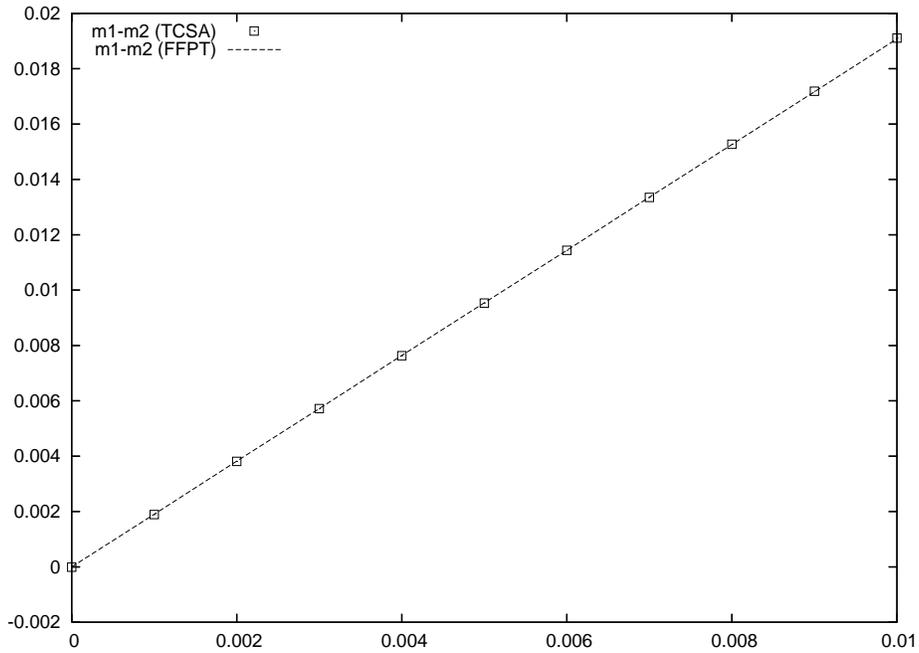}}\end{center}

\caption{\label{cap:tcsa_ffpt} Comparing masses measured from TCSA and predicted
using FFPT for $\beta=\frac{\sqrt{4\pi}}{2.5}\;,\;\delta=0$. All
data are in units of the unperturbed soliton mass $M$. $m1$ and
$m2$ denote the two masses $m_{1}^{(1,2)}$, respectively. The error
bars of TCSA data are too small to visualize.}
\end{figure}

At this value of $\beta$ there are\[
\left[\frac{8\pi}{\beta^{2}}\right]-1=10\]
breathers in the $\lambda=0$ spectrum which shows that this model
is deeply in the semiclassical regime. The FFPT formula (\ref{eq:ffpt_b1})
gives the following prediction for the mass corrections:\begin{equation}
\frac{\delta m_{1}^{(1,2)}}{M}=\pm0.95392713\dots\times\chi\label{eq:d0_ffpt_prediction}\end{equation}
We summarize the measured values of the masses in table \ref{cap:mass_2.5_0}
and plotted them against predictions of form factor perturbation theory
in figure \ref{cap:tcsa_ffpt}. It is obvious that the measured masses
follow well the predictions from form factor perturbation theory,
but there are deviations that grow with the coupling. These are expected
to be higher order effects. However, second order corrections should
cancel from the difference, since sending $\lambda$ to $-\lambda$
exchanges the identity of the two particles \cite{dsg}. In fact,
the mass differences agree with form factor perturbation theory within
the estimated TCSA errors which can be estimated from cut-off dependence
of the energy levels and the flatness of the gap function in the scaling
region. We can also fit the coefficient of (\ref{eq:d0_ffpt_prediction})
against the mass splitting measured in TCSA, for which we find\[
0.95441\pm0.00032\]
The central value differs by only $0.0005$ from (\ref{eq:d0_ffpt_prediction}),
consistent with the $1\sigma$ error range of the fit.

\begin{table}
\begin{center}\begin{tabular}{|c|c|c|c|c|c|c|}
\hline 
$\chi$&
$m_{1}^{(1)}$ &
$m_{1}^{(2)}$ &
$m_{1}^{(1)}$ &
$m_{1}^{(2)}$ &
$m_{1}^{(1)}-m_{1}^{(2)}$&
$m_{1}^{(1)}-m_{1}^{(2)}$ \tabularnewline
&
(TCSA)&
(TCSA)&
(FFPT)&
(FFPT)&
(TCSA)&
(FFPT)\tabularnewline
\hline
\hline 
0&
0.27230&
0.27231&
0.27233&
0.27233&
-0.00001&
0\tabularnewline
\hline 
0.\.{0}01&
0.27325&
0.27136&
0.27328&
0.27138&
0.00019&
0.00020\tabularnewline
\hline 
0.002&
0.27420&
0.27039&
0.27424&
0.27042&
0.00381&
0.00380\tabularnewline
\hline 
0.003&
0.27514&
0.26942&
0.27519&
0.26947&
0.00572&
0.00572\tabularnewline
\hline 
0.004&
0.27610&
0.26847&
0.27615&
0.26851&
0.00763&
0.00764\tabularnewline
\hline 
0.005&
0.27702&
0.26749&
0.27710&
0.26756&
0.00953&
0.00954\tabularnewline
\hline 
0.006&
0.27796&
0.26652&
0.27805&
0.26661&
0.01144&
0.01144\tabularnewline
\hline
0.007&
0.27888&
0.26553&
0.27901&
0.26565&
0.01335&
0.01336\tabularnewline
\hline
0.008&
0.27982&
0.26455&
0.27996&
0.26470&
0.01527&
0.01526\tabularnewline
\hline
\end{tabular}\end{center}

\caption{\label{cap:mass_2.5_0} Comparing masses measured from TCSA and predicted
using FFPT for $\beta=\frac{\sqrt{4\pi}}{2.5}\;,\;\delta=0$. All
data are in units of the unperturbed soliton mass $M$. TCSA mass
measurements have an estimated error of order $\pm0.00002$.}
\end{table}

It is also interesting to examine the mass correction in the classical
limit. Calculating the curvature of the field potential around its
minima at $\varphi=0$ and $\varphi=\frac{2\pi}{\beta}$ gives\begin{equation}
\frac{\delta m_{1}^{(1,2)}}{M_{0}}=\pm\chi_{0}\qquad,\qquad\chi_{0}=\frac{\lambda_{0}}{M_{0}^{2}}\label{eq:d0_classical_prediction}\end{equation}
We can see that quantum corrections to the coefficient are of the
order $0.05$, and the TCSA method is able to measure them with one
percent accuracy. It is also useful to remark that the semiclassical
expansion of quantum corrections to mass ratios goes in the parameter\[
\frac{\beta^{2}}{8\pi}\]
(as can be seen e.g. by expanding the unperturbed sine-Gordon breather/soliton
mass ratio (\ref{eq:br_masses}) for small $\beta$) and for this
particular value of $\beta$ we have\[
\frac{\beta^{2}}{8\pi}=0.08\]
which explains why the corrections to the classical limit are so small.

\subsection{$\delta=\pi/2$: the acid test}

In this case there is a symmetry\begin{equation}
\varphi\;\rightarrow\;\frac{2\pi}{\beta}-\varphi\label{eq:Z2_d2}\end{equation}
which we used for projection. We again performed runs with and without
a projection to confirm that the program was functioning correctly.
In this case, both the even and the odd sector contains a vacuum and
a one-particle state.

For the value of $\beta$ used in the previous section, we obtain
the masses listed in table (\ref{cap:mass_2.5_d2}). At the same truncation
level ($18$) the dimension of the even sector is $9334$ while that
of the odd is $9319$ as before, but the state vectors assigned to
these sectors are different since now we use (\ref{eq:Z2_d2}) instead
of (\ref{eq:Z2_d0}) for the projection. In the unperturbed sine-Gordon
model ($\lambda=0$) both projections can be done and the spectra
compared to check the validity of the program, and we found a perfect
match within the precision of floating point arithmetics.

\begin{figure}
\begin{center}\includegraphics{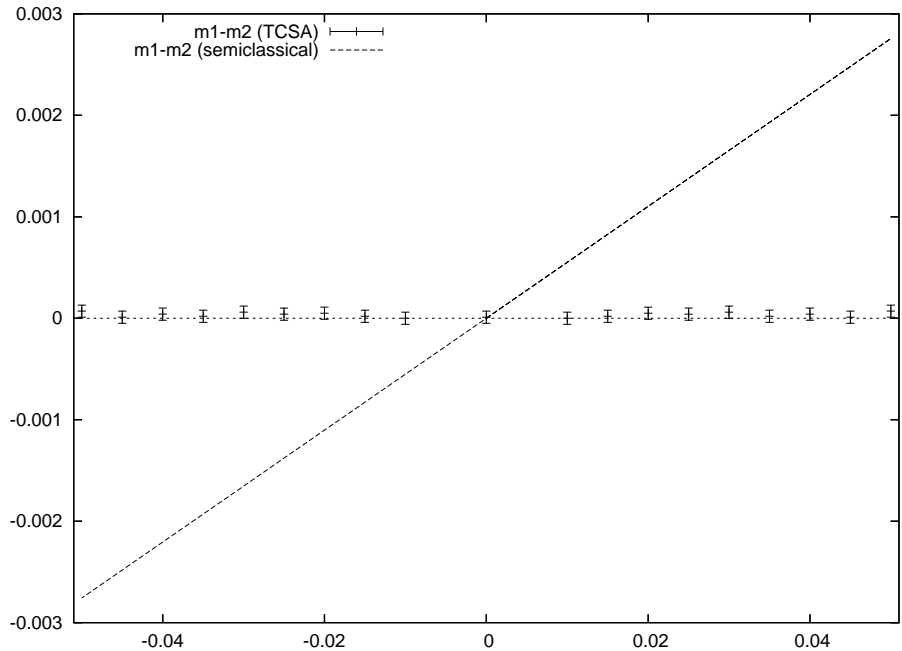}\end{center}

\caption{\label{cap:mass_split25} The mass splitting measured from TCSA compared
to the semiclassical prediction (\ref{eq:mussardo_prediction}) at
$\beta=\frac{\sqrt{4\pi}}{2.5}\;,\;\delta=\frac{\pi}{2}$. The error
bars indicate the estimated magnitude of truncation errors.}
\end{figure}

The spectrum does not change if we change the sign of $\chi$ in either
sectors, so the masses $m_{1}^{\pm}$ are even functions of $\lambda$.
This fact alone already precludes the existence of the linear correction
(\ref{eq:mussardo_prediction}), but we shall go through a detailed
analysis in order to take into account the numerical uncertainties
of the TCSA mass determination as well. We can also plot the measured
mass difference and the prediction from (\ref{eq:mussardo_prediction})
as in figure \ref{cap:mass_split25}, and the mismatch is obvious
from the plot.

\begin{table}
\begin{center}\begin{tabular}{|c|c|c|c|c|}
\hline 
$\chi$&
$m_{1}^{+}$&
$m_{1}^{-}$&
$m_{1}^{+}-m_{1}^{-}$ (TCSA)&
$m_{1}^{+}-m_{1}^{-}$ (\ref{eq:mussardo_prediction})\tabularnewline
\hline
\hline 
0&
0.27230&
0.27231&
-0.00001&
0\tabularnewline
\hline 
$\pm$0.010&
0.27154&
0.27154&
0&
0.00033\tabularnewline
\hline 
$\pm$0.015&
0.27059&
0.27057&
0.00002&
0.00050\tabularnewline
\hline 
$\pm$0.020&
0.26925&
0.26925&
0&
0.00066\tabularnewline
\hline 
$\pm$0.025&
0.26750&
0.26746&
-0.00004&
0.00083\tabularnewline
\hline 
$\pm$0.030&
0.26536&
0.26530&
0.00006&
0.00099\tabularnewline
\hline 
$\pm$0.035&
0.26276&
0.26274&
0.00002&
0.00116\tabularnewline
\hline 
$\pm$0.040&
0.25975&
0.25971&
0.00004&
0.00132\tabularnewline
\hline 
$\pm$0.045&
0.25625&
0.25624&
0.00001&
0.00149\tabularnewline
\hline 
$\pm$0.050&
0.25236&
0.25229&
0.00007&
0.00165\tabularnewline
\hline
\end{tabular}\end{center}

\caption{\label{cap:mass_2.5_d2} Masses measured from TCSA for $\beta=\frac{\sqrt{4\pi}}{2.5}\;,\;\delta=\frac{\pi}{2}$
in units of the unperturbed soliton mass $M$. TCSA mass measurements
have an estimated error of $\pm0.00002$. We also list the mass splittings
measured from TCSA and compare it to the predictions (\ref{eq:mussardo_prediction}). }
\end{table}

We can also fit a function of the form\[
m_{1}^{\pm}=m_{1}+a_{\pm}\chi+b_{\pm}\chi^{2}+c_{\pm}\chi^{3}+d_{\pm}\chi^{4}\]
to the measured masses in the $\chi\in[-0.05\;\dots\;0.05]$ range.
The fit results are\[
\begin{array}{lll}
a_{+}=0\pm0.00010 &  & a_{-}=0\pm0.000054\\
b_{+}=-7.606\pm0.018 &  & b_{-}=-7.625\pm0.011\\
c_{+}=0\pm0.054 &  & c_{-}=0\pm0.033\\
d_{+}=-152.4\pm6.9 &  & d_{-}=-149.5\pm4.4\end{array}\]
This must be compared to the prediction (\ref{eq:mussardo_prediction})
which gives $a_{\pm}=\pm0.0165367$. We see that this value is two
orders of magnitude larger than the uncertainty of the fit. Since
the model in question is very deeply in the semiclassical value, even
higher order quantum corrections cannot come to the rescue (we have
seen in the previous section that they are at the level of a few percent).
The conclusion is that the prediction (\ref{eq:mussardo_prediction})
is not valid to the order $\beta^{4}$ which makes the conclusions
of \cite{semicl} based on these results unfounded (an analysis of
the theoretical arguments in that paper is also given in the following
section).

\begin{figure}
\begin{center}\includegraphics{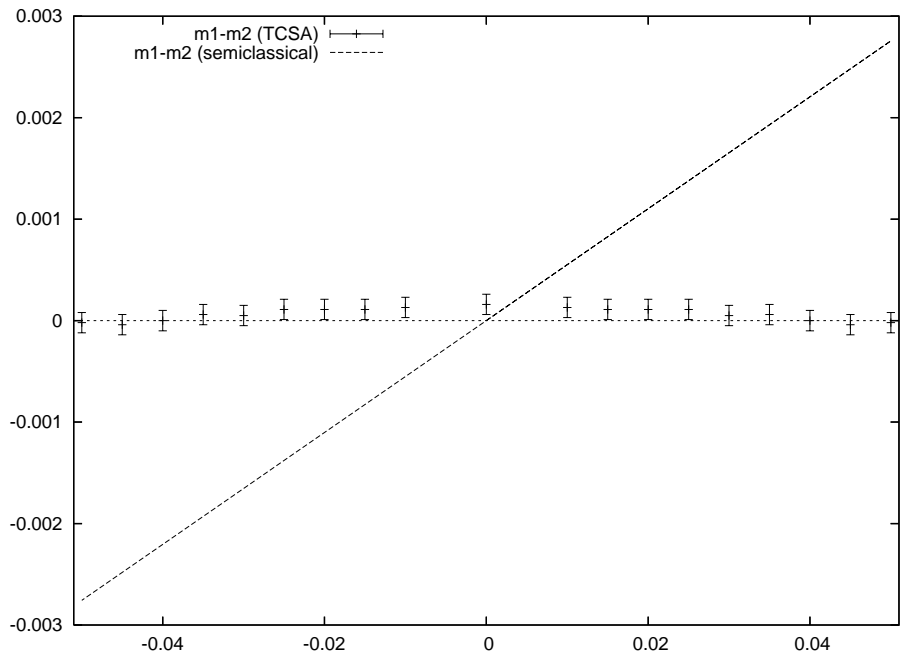}\end{center}

\caption{\label{cap:mass_split22} The mass splitting measured from TCSA compared
to the semiclassical prediction (\ref{eq:mussardo_prediction}) at
$\beta=\frac{\sqrt{4\pi}}{2.5}\;,\;\delta=\frac{\pi}{2}$. The error
bars indicate the estimated magnitude of truncation errors.}
\end{figure}

We also analyzed the mass spectrum for $\beta=\frac{\sqrt{4\pi}}{2.2}$.
This model has $8$ breathers, so it is also very much in the semiclassical
regime ($\beta^{2}/8\pi=0.10331\dots)$. Comparing FFPT with classical
mass corrections again gives an estimate for quantum corrections at
the few percent level. However, higher $\beta$ means that the accuracy
of TCSA is necessarily somewhat lower, but is still enough for our
purposes. At truncation level $18$, we found $7963$ states in the
even and $7978$ states in the odd sector. The measured masses are
summarized in table \ref{cap:mass_2.2_d2}, and the mass splitting
is plotted in figure \ref{cap:mass_split22}.

\begin{table}
\begin{center}\begin{tabular}{|c|c|c|c|c|}
\hline 
$\chi$&
$m_{1}^{+}$&
$m_{1}^{-}$&
$m_{1}^{+}-m_{1}^{-}$ (TCSA)&
$m_{1}^{+}-m_{1}^{-}$ (\ref{eq:mussardo_prediction})\tabularnewline
\hline
\hline 
0&
0.36002&
0.35986&
0.00016&
0\tabularnewline
\hline 
0.010&
0.35940&
0.35927&
0.00013&
0.00055\tabularnewline
\hline 
0.015&
0.35864&
0.35853&
0.00011&
0.00083\tabularnewline
\hline 
0.020&
0.35758&
0.35747&
0.00011&
0.00110\tabularnewline
\hline 
0.025&
0.35622&
0.35611&
0.00011&
0.00137\tabularnewline
\hline 
0.030&
0.35449&
0.35444&
0.00005&
0.00165\tabularnewline
\hline 
0.035&
0.35248&
0.35242&
0.00006&
0.00193\tabularnewline
\hline 
0.040&
0.35011&
0.35011&
0&
0.00220\tabularnewline
\hline 
0.045&
0.34747&
0.34751&
-0.00004&
0.00248\tabularnewline
\hline 
0.050&
0.34443&
0.34445&
-0.00002&
0.00276\tabularnewline
\hline
\end{tabular}\end{center}

\caption{\label{cap:mass_2.2_d2} Masses measured from TCSA for $\beta=\frac{\sqrt{4\pi}}{2.2}\;,\;\delta=\frac{\pi}{2}$
in units of the unperturbed soliton mass $M$. TCSA mass measurements
have an estimated error of order $\pm0.00005$. We also list the mass
splittings measured from TCSA and compare it to the predictions (\ref{eq:mussardo_prediction}). }
\end{table}

We can again perform a fit in the range $\chi\in[-0.05\;\dots\;0.05]$
with the results \[
\begin{array}{lll}
a_{+}=0\pm0.00008 &  & a_{-}=0\pm0.0002\\
b_{+}=-6.076\pm0.017 &  & b_{-}=-5.966\pm0.021\\
c_{+}=0\pm0.053 &  & c_{-}=0\pm0.11\\
d_{+}=-63.3\pm6.5 &  & d_{-}=-77.3\pm8.6\end{array}\]
The prediction (\ref{eq:mussardo_prediction}) gives \[
a_{\pm}=\pm0.0275751\]
which is again ruled out by two orders of magnitude.

\section{Theoretical arguments}

In the previous section we demonstrated that TCSA data are strongly
incompatible with the semiclassical analysis of the mass spectrum
in \cite{semicl}, while they are in full accordance with the analysis
of our earlier paper \cite{dsg} based on form factor perturbation
theory.

\subsection{Issues concerning the semiclassical form factor method}

Considering that the prediction (\ref{eq:mussardo_d2_orig}) is only
semiclassical, one may hope that higher order corrections can adjust
it to be compatible to the numerical data. This hope is unfounded,
however, since we have seen that quantum corrections can be estimated
to be at most a few percent at the values of $\beta$ we considered.
Therefore we must conclude that (\ref{eq:mussardo_d2_orig}) is not
correct even to its lowest order.

This is not surprising in the light of the fact that the leading order
term (\ref{eq:mussardo_leading}) is of order $\beta^{4}$. The semiclassical
computation presented in \cite{semicl} is simply not valid to this
order. Quantum corrections to soliton form factor and soliton mass
should be taken into account before extracting the breather mass from
the pole of the form factor. To order $\beta^{2}$, however, (\ref{eq:mussardo_leading})
is in accord with form factor perturbation theory that predicts no
linear correction to the masses. 

We would also like to discuss briefly the mass corrections predicted
in \cite{semicl} for the case $\delta=0$. They obtain\[
\delta m_{n}^{(\pm)}=\pm\frac{\beta\lambda_{0}}{8\sqrt{\mu_{0}}}\left[\left(1-\log\frac{\lambda_{0}}{16\mu_{0}}\right)\frac{32}{\beta^{2}}\sin\left(n\frac{\beta^{2}}{32}\right)+n\log\frac{\lambda_{0}}{16\mu_{0}}\cos\left(n\frac{\beta^{2}}{32}\right)\right]+O\left(\lambda_{0}^{2}\right)\]
Expanding in $\beta$ \begin{equation}
\frac{\delta m_{n}^{\pm}}{M_{0}}=\pm\left(n+\beta^{4}\frac{n^{3}}{6144}\left(1-\log\frac{\lambda_{0}^{2}}{256\mu_{0}^{2}}\right)+O\left(\beta^{6}\right)\right)\frac{\lambda_{0}}{M_{0}^{2}}+O\left(\lambda_{0}^{2}\right)\label{eq:mussardo_d0_expanded}\end{equation}
The classical limit \[
\frac{\delta m_{n}^{\pm}}{M_{0}}=\pm n\frac{\lambda_{0}}{M_{0}^{2}}+O\left(\lambda_{0}^{2}\right)\]
is correctly reproduced by this formula. However, the $\beta^{4}$
term cannot be trusted because of the same reasons as above. 

Expansion of the FFPT mass correction (\ref{eq:d0_ffpt_prediction})
produces terms of all orders in $\beta^{2}$. It is obvious that a
$\beta^{2}$ term is also generated from (\ref{eq:mussardo_d0_expanded})
when we replace the classical parameters $\lambda_{0}$ and $M_{0}$
by the PCFT coupling $\lambda$ and the quantum soliton mass $M$
since the coupling and the mass are redefined at each order in the
$\beta^{2}$ expansion. For this reason alone the terms in (\ref{eq:mussardo_d0_expanded})
that are higher order in $\beta$ cannot be compared directly to (\ref{eq:d0_ffpt_prediction}). 

However, there is a deeper problem with (\ref{eq:mussardo_d0_expanded}):
the presence of the nonanalytic (logarithmic) dependence in $\lambda_{0}$
would suggest that although the masses have the correct limit when
$\lambda_{0}\,\rightarrow\,0$, the $\lambda_{0}\neq0$ case cannot
be treated as a perturbation of the $\lambda_{0}=0$ model.

In contrast, both conformal perturbation theory and form factor perturbation
theory shows a rather different behaviour. Conformal perturbation
theory produces a series expansion in $\lambda$ and $\mu$, which
contains no ultraviolet divergences at any order if $\beta^{2}<4\pi$.
There does not seem to be any way to produce nonanalytic behaviour
in the coupling $\lambda$ in the regime of small $\beta$: even a
partial resummation over $\mu$ would leave us with an analytic expansion
in powers of $\lambda$ despite the fact that it may produce nonanalytic
terms in $\mu$ (due to infrared divergences present in the perturbation
around the conformal field theory describing the massless ultraviolet
fixed point). This partial resummation, on the other hand, is just
expected to be identical to form factor perturbation theory (which
is free of the infrared divergences since it expands around a massive
field theory) which gives a first order correction in $\lambda$ with
no $\log\lambda$ terms, and matches very well with the numerics of
TCSA as shown in subsection 3.1. On the other hand, since the semiclassical
soliton form factor calculation in \cite{semicl} is not self-consistent
to order $\beta^{4}$, it cannot be used to argue for the presence
of $\log\lambda$ corrections. Before attempting any comparison to
TCSA results, one must compute all the relevant loop corrections and
suitably renormalize $\lambda_{0},\; M_{0}$ to $\lambda,\; M$, which
is out of the scope of the present work. 

We remark that even if all of the above is carried out, it could prove
very difficult to provide any conclusive evidence against the existence
of $\lambda\log\lambda$ corrections of order $\beta^{4}$, since
these cannot be separated easily from the already present $\lambda$
corrections, which are of order $\beta^{0}$. However, the agreement
of the measured mass splitting (illustrated in figure \ref{cap:tcsa_ffpt}(b))
with the FFPT result (\ref{eq:d0_ffpt_prediction}) is very suggestive:
the linear coefficient can be determined to a relative accuracy of
$5\times10^{-4}$, and we still have not exhausted all resources to
make the TCSA even more accurate: increasing the cut level is still
a definite possibility (although it increases the required computer
time very fast, with around the $3.2$th power of the dimension of
the Hilbert space), and extrapolating the cutoff dependence of measured
masses to infinite cutoff is another one.

\subsection{FFPT and neutral states in the $\delta=\frac{\pi}{2}$ case}

In the paper \cite{semicl} the authors also produce a scenario in
the $\delta=\frac{\pi}{2}$ case to support the presence of the linear
mass correction term (\ref{eq:mussardo_d2_orig}). Let us recall the
main line of their argument.

It is known that the breathers arise as bound states in the soliton-antisoliton
($s\bar{s}$) channel, with alternating $C$-parity. Namely, the $n$th
breather $B_{n}$ with $n$ odd/even arises as a bound state in a
channel where the $s\bar{s}$ wave-function is odd/even under charge
conjugation (interchanging soliton and antisoliton), respectively.

The authors argue that such a situation is very unique to the sine-Gordon
model and holds only if the non-integrable coupling $\lambda$ is
not present. They claim that there exists another breather state of
{}``wrong'' $C$-parity and the $n$ breather is actually a doublet
$B_{n}^{\pm}$ (the upper index denoting $C$-parity), but the states
of wrong parity $B_{n}^{-(-1)^{n}}$ decouple from the model which
means that the corresponding three-particle couplings have the behaviour\[
f_{s\bar{s}}^{B_{n}^{+}}=0\quad\mathrm{for}\quad n\quad\mathrm{odd}\qquad f_{s\bar{s}}^{B_{n}^{-}}=0\quad\mathrm{for}\quad n\quad\mathrm{even}\]
so that only one pole appears in the $s\bar{s}$ channel for any $n$,
with alternating $C$-parity, since the other residue vanishes, and
therefore the $2\times2$ $s\bar{s}$ scattering matrix degenerates
to a one-dimensional projection at each pole.

They continue by demonstrating that a modified form factor perturbation
theory, adjusted to take into account this degeneracy produces a linear
correction to the breather mass of equal magnitude and opposite sign
to the two breather copies, thus making the picture appear consistent
with their semiclassical result (\ref{eq:mussardo_d2_orig}). For
$k$-folded models where the periodicity of the field is determined
so that vacua are only identified after $k$ periods\[
\varphi\equiv\varphi+k\frac{2\pi}{\beta}\]
they predict $2k$ copies of each breather, twice as much as the $k$-folded
model has, half of which is claimed to decouple from the model.

There are, however, several problems with this scenario. Extensive
investigation of the spectrum of sine-Gordon theory has been carried
out using the TCSA approach. No sign of these extra states was ever
detected, although the Hilbert space was guaranteed to be complete
by taking all states of the ultraviolet conformal field theory into
account (truncated by an upper energy cut, but all the experience
with TCSA shows that this cannot be the reason for missing these states).
Varying $\beta$ down from the free fermion value $\sqrt{4\pi}$ one
can observe directly how the breather states appear. The result can
be stated very simply in the case of the 1-folded model. We can identify
the $s\bar{s}$ states in the spectrum by fitting their finite size
energy dependence using Bethe Ansatz and the exact $S$-matrix \cite{frt2},
and since the $S$-matrix eigenvalue in the odd/even parity channel
is different, we can even identify the parity of the states. When
$\beta$ crosses the threshold corresponding to the appearance of
the $n$ breather, a single $s\bar{s}$ state crosses below the $2M$
limit (where $M$ is the soliton mass), and this is a state of $C$-parity
$(-1)^{n}$: new breather states appear one by one, and always with
the right parity. There is no trace of the extra states, and this
carries over to the $k$-folded case as well: at each threshold value
of $\beta$ we always observe $k$ copies of the appropriate breathers
appearing in the spectrum (cf. \cite{kfold} for some examples of
spectra; we have produced many of them for different values of $\beta$,
of which the published paper naturally contains only a few representative
examples.)

In the above argument, there is one possible loophole: TCSA uses periodic
boundary conditions, and it may be argued that these exclude some
states from the spectrum or produce some identification between states,
leaving only half of the breather states that are actually there in
infinite volume%
\footnote{This issue was raised by G. Mussardo in a private discussion. There
are known examples where the TCSA spectrum does not include all states
of the infinite volume theory because of the selectivity of boundary
conditions, however they are not very useful analogies of double sine-Gordon
theory, similarly to the case of $RSOS_{3}$ discussed in the main
text.%
}. However, even in this case one expects the mass spectrum extracted
from TCSA to display the correct dependence on the couplings, and
so the fact that our numerical results present such a strong case
against any linear term (or, in fact, any odd power) in $\lambda$
remains a very compelling evidence against this scenario. 

The paper \cite{semicl} also brings up an analogy with the theory
$RSOS_{3}$ (the factorized scattering theory of the tricritical Ising
model). The corresponding $S$-matrix has no bound state poles, but
such poles can be introduced by dressing with appropriate CDD factors.
$RSOS_{3}$ has three vacua which can be illustrated by the potential

\begin{center}\includegraphics{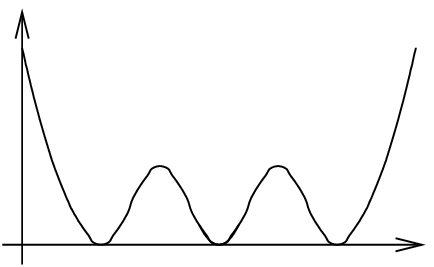}\end{center}

The middle well of these is flanked by two others, and a simple calculation
of the bootstrap spectrum reveals two breather bound states over it.
These can be described as bound states of a kink going from the middle
vacuum to one side and of a kink coming the way back. This seems to
be analogous to the sine-Gordon vacuum structure, which is periodically
infinite, but any chosen vacuum has again two neighbours.

The analogy is flawed, however, because in sine-Gordon theory due
to the periodic vacuum structure the boson is actually compactified
on a circle. In the standard (1-folded) case, the identification is
\[
\varphi\equiv\varphi+\frac{2\pi}{\beta}\]
and therefore all vacua are identified. It does not make much sense
to talk about two vacua flanking a middle one as in the $RSOS_{3}$
case where all three vacua are physically distinct and show up as
$3$ different energy levels in a finite volume (e.g. TCSA) spectrum,
which become exponentially degenerate in the large volume limit. The
$1$-folded sine-Gordon model, on the other hand, has a unique ground
state even in the large volume limit. In the $k$-folded model there
are $k$ different vacua (with the topology of a circle), which is
also manifest in the finite volume spectrum \cite{kfold}. This vacuum
structure is corroborated both by TCSA studies and by the exact description
of the finite volume spectrum using the Destri-de Vega nonlinear integral
equation (NLIE) \cite{DdV} (see also \cite{frt1,frt2,kfold}). It
accounts readily for the $k$ copies of each breather, but there is
no additional doubling in the spectrum: the physics of these vacua
is rather different from that of the $RSOS_{3}$ case.

Our final conclusion is that the arguments given by the authors of
\cite{semicl} for these additional states are not well-founded, while
the numerical data show that the dynamics of the model can be described
very well by the usual form factor perturbation theory applied in
\cite{dsg} without any need for adjustments.

\section{Conclusions}

The numerical analysis has shown with quite large precision that the
conclusions of \cite{semicl} regarding the mass spectrum are untenable,
more precisely the linear mass corrections at $\delta=\frac{\pi}{2}$
predicted by them are ruled out by two orders of magnitude. The main
issue seems to be that they trust the results from the semiclassical
form factor approach to an order at which it is simply not consistent
without performing the appropriate loop corrections as well. We also
considered their theoretical arguments and have shown that they do
not stand up to closer scrutiny. 

In contrast, the numerics is fully consistent with our original conclusions
about the mass spectrum in \cite{dsg}. Obviously, this must not be
taken as a complete verification of these results. It is always possible
that there are effects that the numerics misses because of its finite
precision and otherwise limited scope but which may be relevant for
theoretical understanding. For the present we can only say that under
closer examination the picture presented in \cite{dsg} holds up.
That picture is also consistent with many theoretical expectations
and accumulated experience from the long study of two-dimensional
quantum field theory (the 'folklore'), but certainly this is not a
water-tight proof either. 

We conclude by noting that the double sine-Gordon model is an interesting
quantum field theory in its own right, and we hope that continued
work will shed more light on its workings.

\subsection*{Acknowledgments}

G.T. would like to thank G. Mussardo, Z. Bajnok and L. Palla for discussions.
This work was partially supported by the EC network {}``EUCLID'',
contract number HPRN-CT-2002-00325, and Hungarian research funds OTKA
D42209, T037674 and T043582. G.T. was also supported by a Bolyai J{\'a}nos
Research Fellowship.

\end{document}